\documentclass[amsfonts,amsmath]{PoS}
\usepackage{amsfonts,amsmath}

\newcommand{\psibar}{\ensuremath{\bar\psi}}
\newcommand{\chibar}{\ensuremath{\bar\chi}}
\newcommand{\pbp}{\ensuremath{\psibar\psi}}

\newcommand{\vev}[1]{\ensuremath{\left<#1\right>}}
\newcommand{\vevsub}[1]{\ensuremath{\left<#1\right>_{\mathrm{sub}}}}

\newcommand{\rchi}{\ensuremath{\left ( \frac{r_\chi}{a} \right ) }}
\newcommand{\Real}{\ensuremath{\mathrm{Re}}}

\newcommand{\Trace}{\ensuremath{\mathrm{Tr}}}

\newcommand{\Eq}[1]{Eq.\,(\ref{#1})}

\newcommand{\Fig}[1]{Fig.\,\ref{#1}}
\newcommand{\figs}{Figs.\,}

\newcommand{\Cite}[1]{Ref.\ \cite{#1}}

\newcommand{\ie}{i.\ e.\ }

\newcommand{\links}{\textbf{Left}: }
\newcommand{\mitte}{\textbf{Middle}: }
\newcommand{\rechts}{\textbf{Right}: }

\newcommand{\spbp}{\ensuremath{\sigma^2_{\pbp}}}

\title{\begin{center} Pseudo-Critical Temperature and \\ Thermal Equation of State \\ from $N_f=2$ Twisted Mass Lattice QCD \end{center}}

\ShortTitle{}

\author{{\textbf tmfT Collaboration:}}
\author{\speaker{F. Burger}, M. Kirchner, M. M{\"u}ller-Preussker\\
        Humboldt-Universit\"at zu Berlin, Institut f\"ur Physik, 12489 Berlin, Germany
}
\author{E.-M. Ilgenfritz\\
       Joint Institute for Nuclear Research, VBLHEP, 141980 Dubna, Russia}
\author{M. P. Lombardo \\
       Laboratori Nazionali di Frascati, INFN, 100044 Frascati, Roma, Italy}
\author{O. Philipsen, C. Pinke, L. Zeidlewicz  \\
       Goethe-Universit\"at Frankfurt, Institut f\"ur Theoretische Physik, 60438 Frankfurt am Main, Germany}

\abstract{
We report about the current status of our ongoing study of the chiral limit of 
two-flavor QCD at finite temperature with twisted mass quarks. We estimate the 
pseudo-critical temperature $T_c$ for three values of the pion mass in the range 
of $m_\mathrm{PS} \simeq 300$ and $500 \mathrm{~MeV}$ and discuss different 
chiral scenarios.

Furthermore, we present first preliminary results for the trace anomaly, pressure and energy 
density. We have studied 
several discretizations of Euclidean time
up to $N_\tau=12$ in order to assess the continuum limit of the trace anomaly. 
From its interpolation
we evaluate the pressure and energy density employing the integral method. 
Here, we have focussed
on two pion masses with $m_\mathrm{PS} \simeq 400$ and $700 \mathrm{~MeV}$.   
}

\FullConference{The 30 International Symposium on Lattice Field Theory - Lattice 2012,\\
		June 24-29, 2012\\
		Cairns, Australia}

\begin{document}

\section{Introduction}
The order of the phase transition in the case of two-flavor QCD in the chiral limit remains an 
open question. While with Wilson quarks \cite{AliKhan:2000iz,Bornyakov:2009qh} 
the transition is found to be
compatible with a second order phase transition in 
the universality class of a 3d $O(4)$ spin
model, a first order transition seems to be favoured in 
analyses with staggered fermions at $N_\tau=4$
\cite{Bonati:2009zz,Bonati:2012pe}.

The thermal equation of state (EoS) 
constitutes a relevant ingredient in the hydrodynamic evolution
of the quark-gluon plasma created in heavy-ion experiments. 
It can be determined non-perturbatively in lattice calculations.
In the recent past the EoS has been studied extensively using the staggered type of 
quark discretization, mostly with $N_f=2+1$ flavors at the physical point \cite{karscheos, fodoreos}. 
The much more compute-intensive Wilson-like discretizations
are less investigated however \cite{eos_alikhan,Umeda:2012er}.
In the latter study the fixed scale approach is used as compared to the more traditional fixed $N_\tau$ 
approach.

\section{Lattice Setup}
The lattice setup in our ongoing investigations equals the one employed 
by the European Twisted Mass
Collaboration (ETMC) for their
$N_f=2$ simulations \cite{tmanalysisdetails}. It employs the twisted mass action in terms of 
the twisted fields $\chi  = \exp{(- i \pi \gamma_5 \tau_{3} / 4 )} \psi$
\begin{equation}
S^{\mathrm{tm}}_f[U,\psi,\psibar] = \sum_{x,y} \chibar(x) \left 
 ( \delta_{x,y} -\kappa D_\mathrm{W}(x,y)[U] + 2 i \kappa a \mu \gamma_5\tau_3 \delta_{x,y} \right )
 \chi(y) \;.
\label{tmaction}
\end{equation}
in the quark sector, while the gauge sector is described by the 
tree-level Symanzik improved gauge action
\begin{equation}
S_g^{\mathrm{tlSym}}[U] = \beta  \Big{(} c_0 \sum_{P} \lbrack 1 - \frac{1}{3} \Real \Trace \left ( U_{P} \right ) \rbrack 
 + c_1 \sum_{R} \lbrack  1 - \frac{1}{3} \Real \Trace  \left ( U_{R} \right ) \rbrack \Big{)} \; .
\label{tlsym}
\end{equation}
The latter two sums extend over all possible plaquettes ($P$) and all possible planar
rectangles ($R$), respectively.

\section{Pseudo-Critical Temperatures and Chiral Limit}
\begin{figure}[htb]
{\centering 
  \hfill \includegraphics[height=4.4cm]{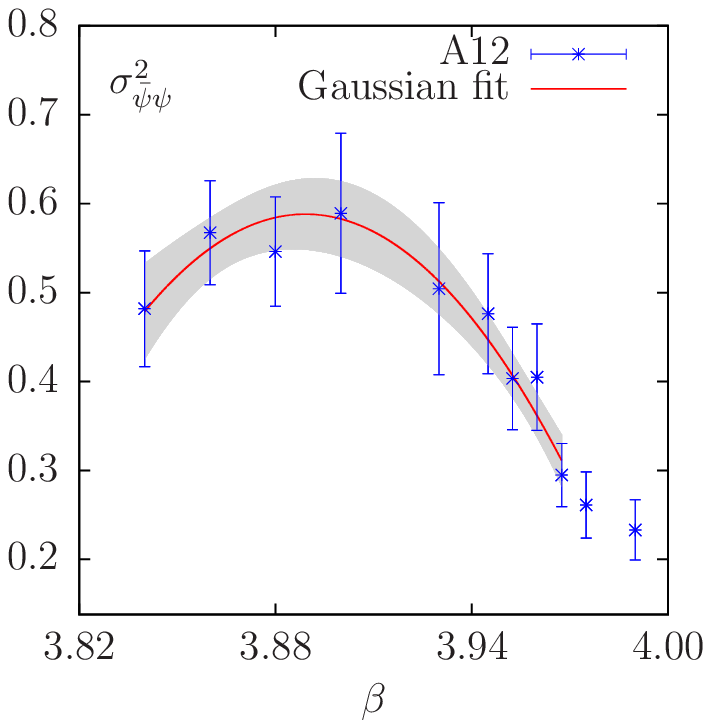} \hfill
  \hfill \includegraphics[height=4.4cm]{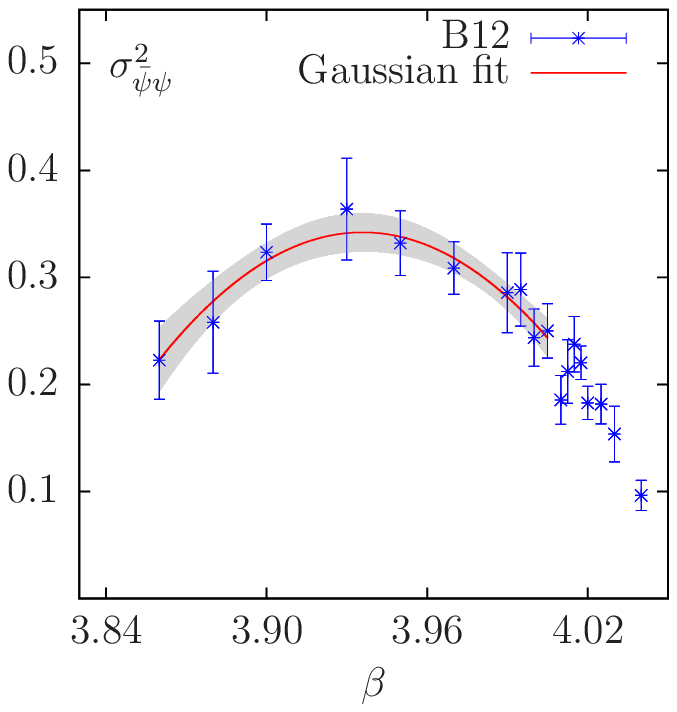} \hfill 
  \hfill \includegraphics[height=4.4cm]{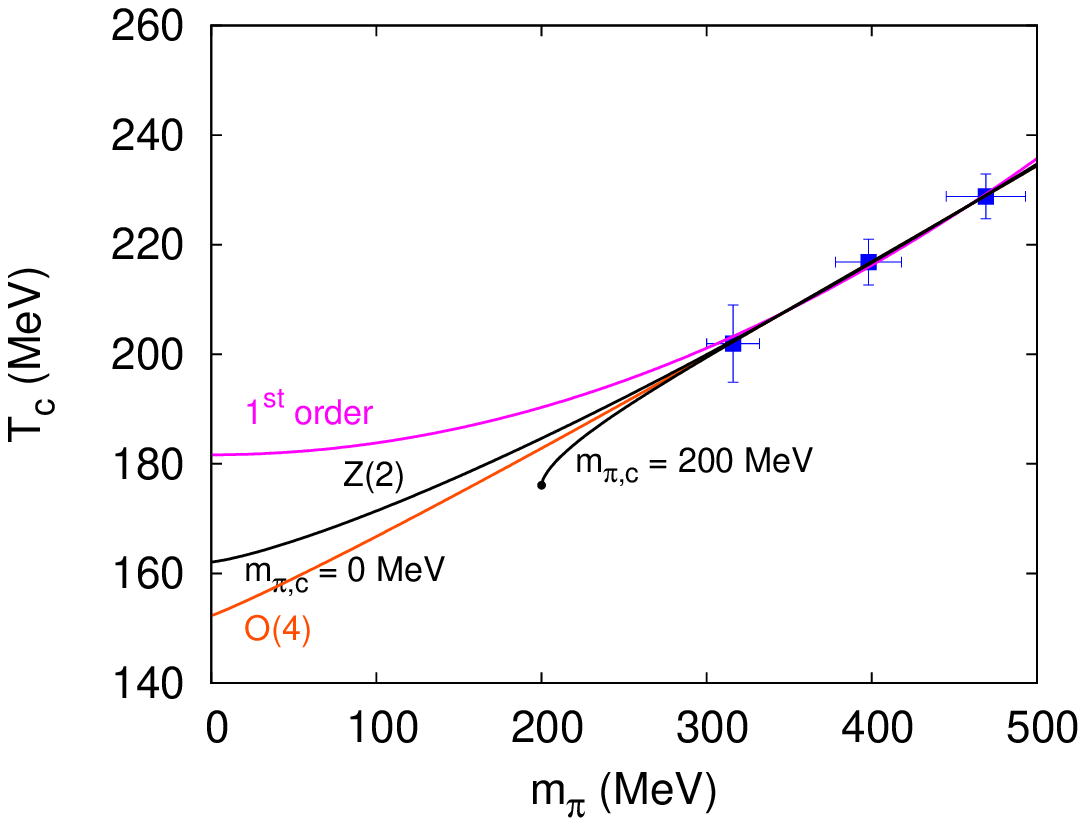} \hfill
}
 \caption{Determination of $T_c$ from $\spbp$ using a Gaussian fit to the maximum for the A12 and B12 ensembles. 
 Comparison of different scenarios for the chiral limit of $T_c$.
}
\label{fig_Tc}
\end{figure}

For the present study of the chiral limit we rely on simulations with $N_\tau=12$ 
at pion masses $m_{\mathrm{PS}}\simeq 320~\mathrm{MeV}$,
$400~\mathrm{MeV}$ and  
$470~\mathrm{MeV}$ that have been 
analyzed in \Cite{Burger:2011zc}
(for historical reasons we call these ensembles A12, B12 and C12).
Our determination of the pseudo-critical temperature is based on the measurement of 
the variance of $\pbp$ over the gauge ensemble 
\begin{equation}
\spbp = \frac{V}{T} \left( \vev{(\pbp)^2} - \vev{\pbp}^2 \right)\;.
\label{eq_sigmachiral}
\end{equation}
It corresponds to the disconnected part of the usual chiral susceptibility
and should show a maximum in the region of $T_c$. This is indeed the case for all our 
ensembles and two representative cases are shown in the two left panels of \Fig{fig_Tc}.
From fitting Gaussian function to the data of $\spbp$ around the maxima we infer values 
of the pseudo-critical couplings $\beta_c$ that are converted to a physical value 
of $T_c$ using an interpolation of $a(\beta)$ \cite{Burger:2011zc}.
At leading order in chiral perturbation theory and for a phase transition of second order
the pion mass dependence of $T_c$
is expected to be given as 
\begin{equation}
 T_c(m_\pi)  =  T_c(0) + A \ m_\pi^{2/(\tilde \beta \delta)} \;,
  \label{eq_Tcfit}
\end{equation}
where $T_c(0)$ is the critical temperature in the chiral limit and $\tilde \beta$ and $\delta$ are 
critical exponents corresponding to the universality class of second order phase transition.
We have restricted ourselves to the chiral scenarios discussed in \Cite{Burger:2011zc} including 
a first order scenario as well as the $O(4)$ and $Z(2)$ second order scenarios, for the latter assuming 
a second order endpoint located at $m_{\pi,c} = 0~\mathrm{MeV}$
or alternatively at $m_{\pi,c} = 200~\mathrm{MeV}$.
The result of fits of \Eq{eq_Tcfit} to our data is shown in the 
right most panel of \Fig{fig_Tc}. As the fitted curves 
are all describing the given data 
quite well we conclude 
that the present set of pion mass values can not discriminate 
among the different chiral scenarios that have been studied. 
For the $O(4)$ model the fit prefers a value of 
$T_c(0) = 152~(26) \mathrm{~MeV} $.

\section{The Trace Anomaly}
For the EoS we concentrate on one of the values of pion masses used in the chiral limit study above,
namely the one corresponding to
$m_{\mathrm{PS}}\simeq 400~\mathrm{MeV}$. 
We have added at the same pion mass 
additional runs at smaller
$N_\tau=4, ~ 6 \mathrm{~and~} 8$ 
(henceforth denoted by B4, B6, B8 $\ldots$). Apart from the latter, for which $N_\sigma=28$ 
has been chosen, all lattices have a spatial extent $N_\sigma=32$. Moreover, further 
ensembles at $m_{\mathrm{PS}}\simeq 700~\mathrm{MeV}$ (further on referred to as the D mass) 
were generated with sizes  
$N_\sigma^3 \times N_\tau = 24^3\times 10,~ 20^3\times 8 \mathrm{~and~} 16^3 \times 6$
(referred to as D10, D8 and D6).

The direct evaluation of pressure $p= T \left. \frac{\partial \ln Z}{\partial V}\right |_{T}$ 
and energy density $\epsilon = \left. \frac{T}{V} \frac{\partial \ln Z}{\partial \ln T}\right |_{V}$ 
from derivatives of the partition function $Z$ is problematic given the lattice spacing dependence of 
both the temperature $T=1/(N_\tau a)$ and the volume $V=N_\sigma^3 a^3$. 
The by now standard approach is the use of the integral method to calculate 
the pressure as a temperature integral of the total derivative of the partition
function with respect to the lattice spacing,
the so called trace anomaly:
\begin{equation}
  \begin{split}
    \frac{I}{T^4} &=  \frac{\epsilon - 3 p}{T^4} = 
                    - \frac{T}{V T^4}\vevsub{ \frac{d \ln Z}{d \ln a} } \\ 
                 &= N_\tau^4 B_\beta \frac{1}{N_\sigma^3 N_\tau} \Bigg \{ \frac{c_0}{3} \vevsub{\Real \Trace \sum_P  U_P} + 
       \frac{c_1}{3} \vevsub{\Real \Trace \sum_R U_R} + B_\kappa  
        \vevsub{\bar \chi D_\mathrm{W}[U] \chi} \\
     &\hphantom{N_\tau^4 B_\beta \frac{1}{N_\sigma^3 N_\tau} \Bigg \{ } - \left [ 2 (a \mu)  B_\kappa  + 2 \kappa_c (a \mu) B_\mu  \right ]
       \vevsub{\bar \chi i \gamma_5 \tau^3 \chi} \Bigg \} \;.   
       \end{split}
\label{eq_traceanomaly_pre}
\end{equation}
Here, $B_\beta$, $B_\mu$ and $B_\kappa$ are related to the $\beta$-functions,
the derivatives of the bare parameters with respect to the lattice spacing, as follows:
\begin{equation}
  B_\beta =  a \frac{d \beta}{d a} \;, \quad \quad 
  B_\mu = \frac{1}{(a \mu)}\frac{\partial (a \mu)}{\partial \beta} \;, \quad \quad 
  B_\kappa =  \frac{\partial{\kappa_c}}{\partial{\beta}} \;. 
\end{equation}
It is necessary to subtract from each term in above expression
the corresponding $T=0$ vacuum contribution 
$\vevsub{\ldots} \equiv \vev{\ldots}_{T>0} - \vev{\ldots}_{T=0}$ in order to achieve a finite result.
Details of the subtraction on the basis of the available $T=0$ lattice data as well as on the 
evaluation of the $\beta$-functions will be given in the following section. In the left panels 
of \figs \ref{fig_D8em3p} and \ref{fig_B12em3p} we
show the trace anomaly 
for the two cases of pseudoscalar masses under investigation.
In both cases we observe sizeable
lattice artifacts in the height of the maximum and even in the falling edge at 
larger temperatures. Moreover, the precision in case of the smaller mass is not
yet satisfactory, especially at small temperatures.

\section{$\beta$-Functions and $T=0$ Subtraction}
\begin{figure}[htb] 
{\centering
\hfill
\includegraphics[height=3.8cm]{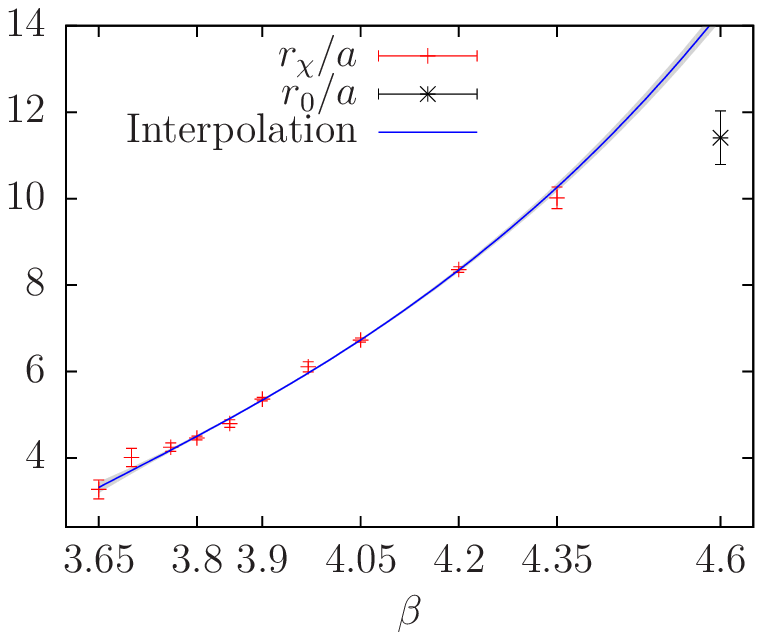}    \hfill
\includegraphics[height=3.8cm]{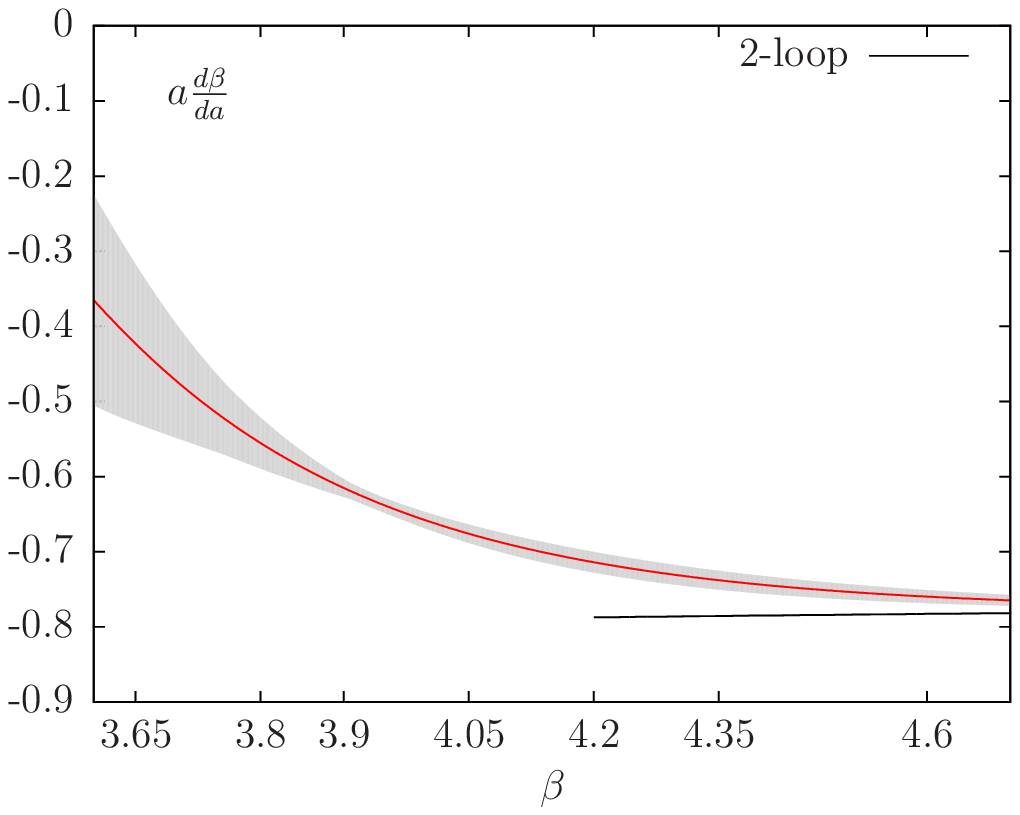}\hfill
\includegraphics[height=3.8cm]{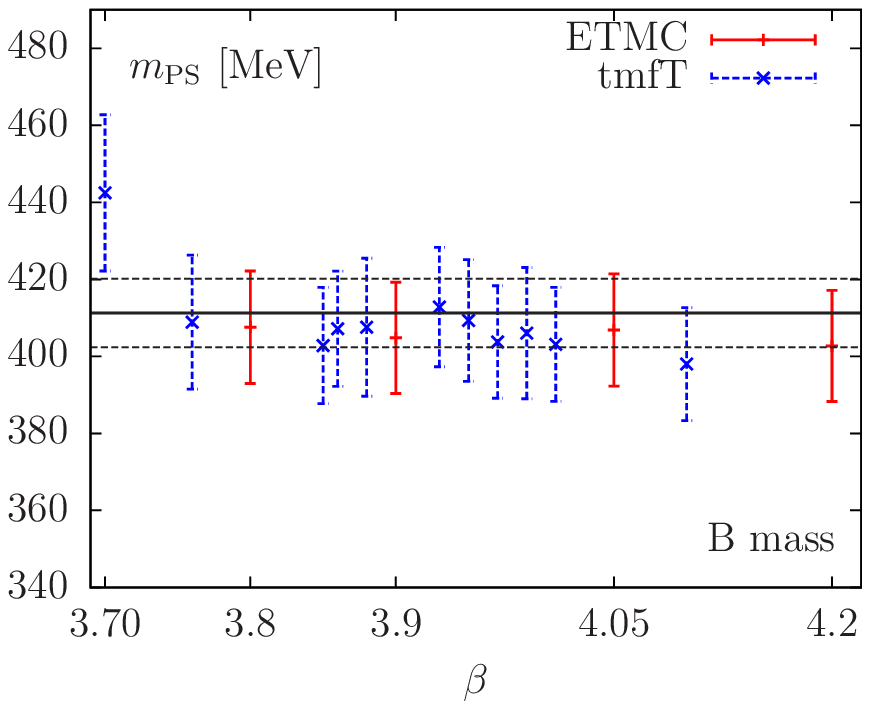}
\hfill
}
 \caption[]{\links Interpolation of 
$r_\chi/a$ in the bare coupling. The point at $\beta = 4.6$ is not obtained 
from a chiral extrapolation and is not included in the fit. 
\mitte The $\beta$-function obtained 
 according to \Eq{eq_betafunction}. We also show the perturbative 2-loop
 expectation at large couplings $ B_\beta \left ( \beta \right ) 
   = - 12 \beta_0 - 72 \frac{\beta_1}{\beta}$. \rechts Pion mass in physical units for the B 
   ensembles together with a constant fit over 
  all data points.}
\label{fig_betafun}
\end{figure}

For evaluating the three $\beta$-functions we consider fits to lattice data of the 
Sommer scale $r_0$ in the chiral limit (denoted by $\rchi$). To this end the correct 
asymptotic behavior is built into the
fit functions explicitly following \Cite{karscheos2007}. For instance we determine $B_\beta$ via the identity 
\begin{equation}
    B_\beta = \left ( a \frac{d \beta}{d a} \right ) = - \rchi \left ( \frac{d \rchi}{d \beta} \right )^{-1} \;,
    \label{eq_betafunction}
\end{equation}
by fitting $r_\chi/a$ to the formula
\begin{equation}
  \rchi (\beta) = 
  \frac{1 + n_0 R(\beta)^2}{d_0 \left ( a_{\mathrm{2L}}(\beta) + d_1 R(\beta)^2 \right ) },
  \quad  \quad 
  R(\beta) = \frac{a_{\mathrm{2L}}(\beta)}{a_{\mathrm{2L}}(\beta_{\mathrm{sub}})} \;.
 \label{eq_betafunfit}
\end{equation}
The ratio $R(\beta)$ is defined in terms of the known
two-loop perturbative formula $a_{\mathrm{2L}}(\beta)$ and 
$\beta_{\mathrm{sub}}=3.9$ has been chosen in above formula. 
The three parameter fit of
\Eq{eq_betafunfit} to $\rchi$ (see the left panel of \Fig{fig_betafun}) yields $\chi^2/\mathrm{dof} = 1.2$. 
The thus obtained 
$\beta$-function is shown in the middle panel of \Fig{fig_betafun}.
The interpolation provided by the fit of \Eq{eq_betafunfit} has also
been used to set the scale using the physical value of $r_0=0.420(15) \mathrm{~fm}$
by ETMC \cite{tmlightmeson}.

The second $\beta$-function associated with the mass is evaluated from a similar identity \cite{karscheos2007}:
\begin{equation}
    B_\mu = \frac{1}{(a \mu)} \frac{\partial (a \mu)}{\partial \beta} = 
    B^{-1}_\beta + \frac{1}{r_\chi \mu} \frac{\partial (r_\chi \mu)}{\partial \beta} \;. \\
\end{equation} 
$r_\chi \mu$ as well as its derivative 
are obtained by fitting the following expression to $r_\chi \mu$:
\begin{equation}
  r_\chi \mu = \left ( \frac{12 \beta_0}{\beta} \right )^{\gamma_0/2\beta_0} P(\beta), \quad \quad 
  P(\beta) = a_\mu \left ( 1 + b_\mu R(\beta)^2 \right ) \;,
\label{eq_mufunctionfit}
\end{equation}
where $\beta_0=(11-2 N_f/3)/(4 \pi)^2$ and $\gamma_0=1/(2 \pi^2)$. 
The third and remaining $\beta$-function involving $B_\kappa$ is calculated in the 
most straight-forward manner from an explicit 
derivative of $\kappa_c$ with respect to $\beta$
using the Pad\'{e} interpolation of \Cite{Burger:2010ag}. 

In order to obtain $\vevsub{\ldots}$, \ie to 
subtract the $T=0$ expectation values, 
we have used all available 
lattice data from ETMC. For these it has been necessary to interpolate in $a \mu$ using
spline functions to match with
the simulated bare mass at $T > 0$. Further additional $T=0$ runs have been simulated 
in order to perform the subtraction more reliably.  However, not all simulation points at finite
temperature are supplemented with an associated $T=0$ simulation. Thus, we have performed an 
interpolation in $\beta$ using a polynomial ansatz of fifth order. For the plaquette, the
rectangle and the Wilson hopping term $D_\mathrm{W}$ we have obtained values for
$\chi^2$ per degree of freedom of $2.7$, $2.3$ and $2.9$, respectively.
The remaining term, for which no fit of reasonable quality could be obtained 
using this ansatz, has been interpolated using splines.
For the D ensembles with larger mass, sufficient $T=0$ data newly generated is available.
Hence, no 
interpolations in the bare coupling and only few interpolations in the bare mass had to be 
done. 

\section{Pressure and Energy Density}
The evaluation of the pressure from the integral technique proceeds by 
integrating the identity $\frac{I}{T^4} = T \frac{\partial}{\partial T} \left (  \frac{p}{T^4}  \right )$ 
in temperature along the line of constant physics (LCP):
\begin{equation}
  \frac{p}{T^4} - \frac{p_0}{T_0^4} = \left . \int_{T_0}^{T} d\tau \frac{\epsilon - 3 p}{\tau^5} \right |_\mathrm{LCP} \;.
\label{eq_pressureintegral}
\end{equation}
We define the LCP in terms of the pion mass  in physical units, 
which for the smaller mass run is shown in the right panel of \Fig{fig_betafun}.
 As can be seen it is constant within errors. For the larger mass, however, we 
observe a systematic rise of  $m_{\mathrm{PS}}$ towards larger coupling 
which amounts to a violation of the LCP condition on the level of $10 \%$.

\begin{figure}[htb] 
{\centering
\hfill
\includegraphics[height=3.8cm]{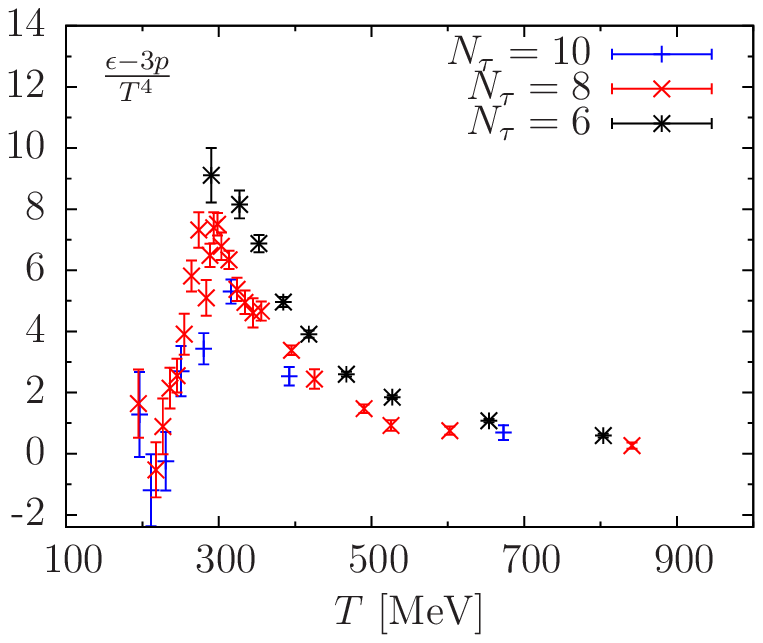} \hfill
\includegraphics[height=3.8cm]{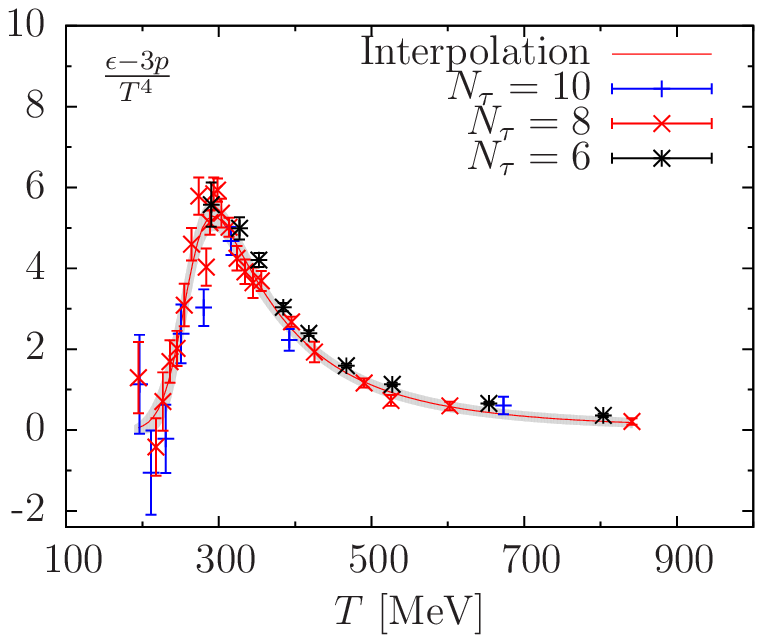}\hfill
\includegraphics[height=3.8cm]{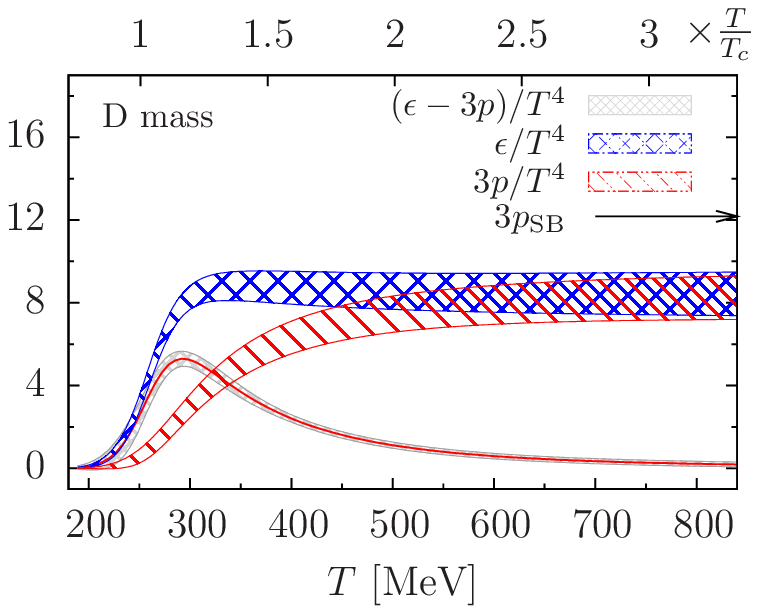}\hfill
}
\caption[EoS for the B ensembles]{
\links The trace anomaly for the D ensembles (see text) obtained for different 
values of the temporal extent $N_\tau$. 
\mitte The trace anomaly after tree-level correction with a fit of 
\Eq{eq_em3p_interpol}. \rechts  Preliminary results for the pressure and 
the energy density.}
\label{fig_D8em3p}
\end{figure}

\begin{figure}[htb] 
{\centering
\hfill
\includegraphics[height=3.8cm]{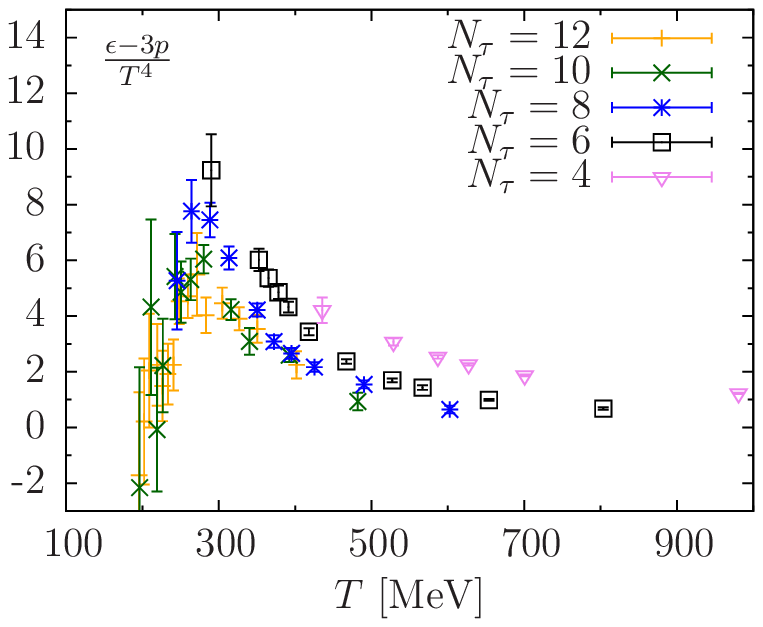} \hfill
\includegraphics[height=3.9cm]{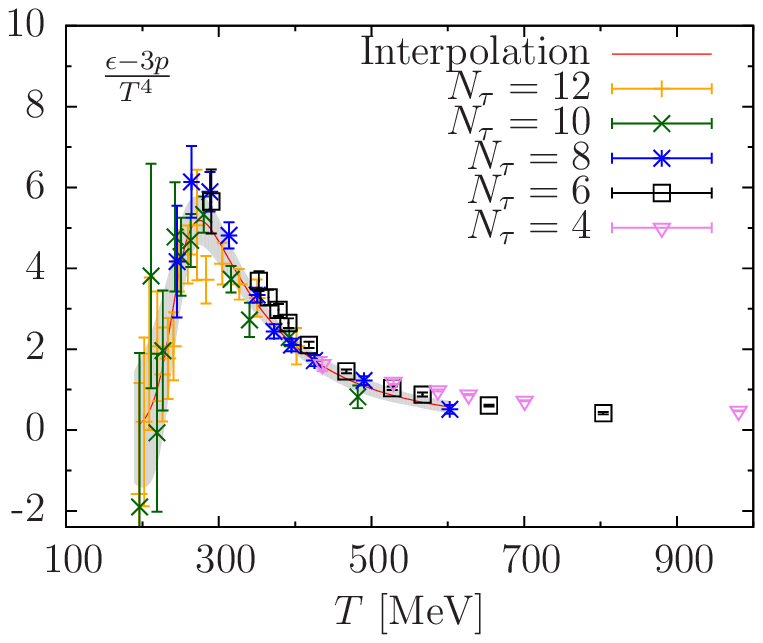} \hfill
\includegraphics[height=3.8cm]{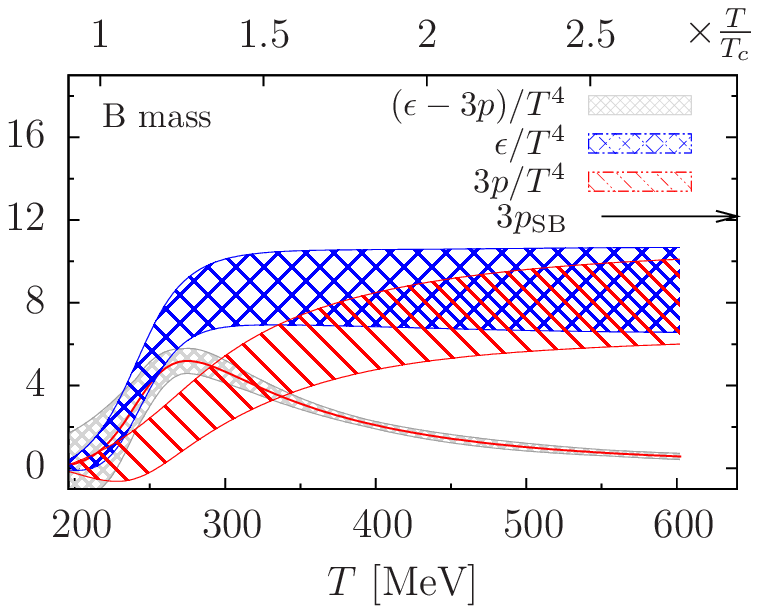}\hfill
}
\caption[EoS for the B ensembles]{The same as in \Fig{fig_D8em3p} but for the B ensembles (see text).}
\label{fig_B12em3p}
\end{figure}

We perform the integration \Eq{eq_pressureintegral} by 
fitting the available lattice data of $\frac{I}{T^4}$ to the ansatz \cite{fodoreos}
\begin{equation}
  \frac{I}{T^4} = \exp{\left ( - h_1 \bar t - h_2 {\bar t}^2 \right )} \cdot
                  \left ( h_0 + 
                         \frac{f_0 \left \{ \tanh{\left ( f_1 \bar t + f_2 \right )} \right \} }
                              {1 + g_1 \bar t + g_2 {\bar t}^2 } 
                   \right ) \; ,
  \label{eq_em3p_interpol}
\end{equation}
where $\bar t = T/T_0$ and $T_0$ is a free parameter in the fit. 
For the fit we use the tree-level corrected data of the trace anomaly that we obtain by normalizing
it with the lattice-to-continuum ratio of the Stefan-Boltzmann pressure 
in the free limit $p_{\mathrm{SB}}^L/p_{\mathrm{SB}}$\footnote{We use  
$p_{\mathrm{SB}}^L/p_{\mathrm{SB}}= 2.586, 1.634, 1.265,  1.134,  1.084$ 
for $N_\tau =	4, 6, 8, 10, 12$, respectively
as obtained in \Cite{Philipsen:2008gq}. The dependence on the mass ($\mu T$) is
found to be very mild and below 1 \% such that the same correction factor is used 
for all temperatures.} following \Cite{fodoreos}. We check the
validity of this approach by comparing the continuum limit values as 
obtained from the corrected as well 
as uncorrected data for various temperatures and find compatible results
in the majority of cases. As can be observed from \figs \ref{fig_D8em3p} and \ref{fig_B12em3p}, where the thus corrected 
trace anomaly is shown for the various available $N_\tau$, the correction is efficient and overlays the data
from different $N_\tau$.

In order to account for the large errors at small
temperatures we perform fits of \Eq{eq_em3p_interpol} to the upper and 
lower 1-$\sigma$ deviations and keep the resulting
difference as the error of the interpolation. For the B (D) ensembles we 
have fitted data from $N_\tau= 8,10$ and $12$ ($N_\tau = 8$ and $10$) 
simultaneously and obtain acceptable fits in both cases.
We subsequently integrate the interpolation curve
numerically in temperature. The integration constant $p_0$ in \Eq{eq_pressureintegral} has been
set to zero in the present evaluation.
The (yet preliminary) results for the pressure and energy density are shown in the right 
panels of \figs \ref{fig_B12em3p} and \ref{fig_D8em3p}. 

\section{Conclusions}
We have presented results for the mass dependence of the pseudo-critical temperature 
for several small values of the pion mass in the range of 
$m_{\mathrm{PS}}\simeq 320~\mathrm{MeV}$ and $m_{\mathrm{PS}}\simeq 470~\mathrm{MeV}$
in a setup with $N_f=2$ Wilson twisted mass quarks at $N_\tau=12$. The comparison of 
different scenarios in the chiral limit is so far inconclusive at the present masses.
Further, we have
presented first, yet preliminary, results of our ongoing project
aiming at the determination of the EoS. The trace anomaly has 
been computed for two values of the pseudoscalar mass of about
$400$ and $700 \mathrm{~MeV}$ and has been tree-level corrected. 
The pressure has been calculated from the 
integral method using a smooth interpolation formula fitted to
the corrected trace anomaly.

\section*{Acknowledgements}
We are grateful to the HLRN supercomputing centers Berlin and Hannover 
as well as the LOEWE-CSC of Goethe-Universit\"at Frankfurt
for providing computing resources for this project. F.B.~and 
M.M.P.~acknowledge support by DFG GK 1504 and SFB/TR 9, respectively. 
O.P. and C.P. are supported by the Helmholtz                                                                                                            
International Center for FAIR within the LOEWE program of the State of Hesse.

\end{document}